\documentclass[%
reprint,
nofootinbib,
amsmath,amssymb,
aps,
]{revtex4-2}

\usepackage{graphicx}% Include figure files
\usepackage{dcolumn}% Align table columns on decimal point
\usepackage{bm}% bold math
\usepackage{xcolor}
\usepackage{upgreek}
\usepackage{appendix}

\graphicspath{{Figures/}}

\begin{document}

\preprint{APS/123-QED}

%\title{Two-dimensional superconducting quantum interference array arrays with Johnson noise: Transfer function and voltage analysis for arrays with spread in the Josephson junctions. }% Force line breaks with \\
% \title{The effect of critical current spreads on the performance of SQUID arrays using Monte Carlo Integration} % Force line breaks with \\
%\title{Investigating the performance of SQUID arrays with critical current spreads using Monte Carlo integration}
\title{Effect of Josephson junction parameter spread \\ on the performance of %high-temperature
SQUID arrays}
% Transfer function and voltage analysis for 2D arrays with non-identical Josephson junctions.

\author{O. A. Nieves}
\email{oscar.nieves@csiro.au}
\author{M. A. Gal\'i Labarias}
\email{galilabarias.marc@aist.go.jp}
\author{A. C. Keser}
\email{aydin.keser@csiro.au}
\author{K.-H. M\"uller}
\author{E. E. Mitchell}
\affiliation{%
 CSIRO Manufacturing, Lindfield, NSW, Australia.
}%


%  \homepage{http://www.Second.institution.edu/~Charlie.Author}
% \affiliation{
%  Second institution and/or address\\
%  This line break forced% with \\
% }%
% \affiliation{
%  Third institution, the second for Charlie Author
% \affiliation{%
%  Authors' institution and/or address\\
%  This line break forced with \textbackslash\textbackslash
% }%

% \collaboration{CLEO Collaboration}%\noaffiliation

\date{\today}% It is always \today, today,

\begin{abstract}
Josephson junctions based on grain boundaries, such as those made of Yttrium Barium Copper Oxide (YBCO), exhibit inherent parameter spreads in their critical current and normal state resistance. This variation in junction properties leads to a decrease in array performance for magnetic sensing applications. Therefore, we must develop a quantitative understanding of how junction parameter spreads impact arrays with different designs. In this paper, we use numerical simulations to investigate how the ensemble averaged voltage modulation depth $\eta$ of one-dimensional SQUID arrays varies with the statistical spread in the junction parameters. In these calculations for arrays we vary the number of junctions, loop inductance and thermal noise strength. We show that $\eta$ decreases with increasing spread, and that this reduction is accelerated further by the number of junctions and SQUID cell inductance, but is robust to changes in the thermal noise strength.
\end{abstract}

%\pacs{Valid PACS appear hewwwre}% PACS, the Physics and Astronomy
%Use showkeys class option if keyword
\maketitle

%\tableofcontents
%__________________________________________________________________________________________________________________________
\section{\label{sec:Intro}Introduction}
Superconducting quantum interference arrays (SQUIDs) and filters (SQIFs) are used in magnetic sensing applications due to their ability to detect small magnetic fields in the femto-Tesla regime \cite{CLA06, FAG06}. The use of high-$T_c$ materials like Yttrium Barium Copper Oxide (YBCO) has enabled the operation of such arrays at 77 Kelvin, and their sensitivity and noise performance depends on the number of Josephson junctions (JJs), loop geometry and junction electrical properties \cite{MUL21, GAL22a, GAL24, NIE24}. For overdamped JJs like in YBCO, the parameters are the critical current $I_{c,k}$ and normal-state resistance $R_k$, where $k$ denotes the junction number. These parameters can vary within the same array, as has been shown from experimental measurements on single YBCO JJs~\cite{JEN09, MIT10, LAM14, DU14, OHK22}. 

Previous theoretical work has shown that a statistical spread in the junction's critical current in dc-SQUIDs~\cite{TES77, JMUL01}, SQUID arrays~\cite{BER14, GAL23} and SQIF arrays~\cite{WU12, MUL24} decreases array performance, measured from the array's dc voltage-flux response in terms of the modulation depth (also known as peak-to-peak voltage). However, these studies have been limited to specific array designs, without looking at the interplay between junction spread and other design parameters such as the number of junctions $N_p$, screening parameter $\beta_L$ or thermal noise strength $\Gamma$ associated with Johnson noise at finite temperatures $T$.

In this paper, we numerically investigate how the ensemble averaged modulation depth $\Delta\overline{v}$ of one-dimensional SQUID arrays varies with the critical current spread $\sigma$ using Monte Carlo simulation. We focus on arrays with identical square loops, while varying the number $N_p$ of junctions in parallel, the screening parameter $\beta_L$, and the thermal noise strength $\Gamma$. 

%__________________________________________________________________________________________________________________________
\section{Methods}\label{sec:The}
\begin{figure}
	\centering
	\includegraphics[width=0.4\textwidth]{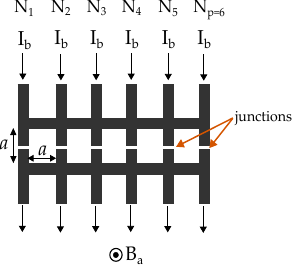}
	\caption{Schematic of a 1D SQUID array with $N_p=6$ using the uniform-biasing scheme: each bias lead receives the same bias current $I_b$. The externally applied magnetic field is denoted as $B_a$.}
	\label{fig:array}
\end{figure}

To model the voltage response of the SQUID arrays we use the previously introduced model~\cite{CYB12, GAL22a} which is based on the RSJ model, and which was later extended to incorporate the statistical spread in the JJ critical currents~\cite{GAL23}. Our model takes mutual inductances fully into account. In this paper, we consider one-dimensional SQUID arrays with $N_p$ junctions connected in parallel, and assume that the arrays are uniformly biased as shown in Fig.~\ref{fig:array}.

In this study, all arrays have identical loops with self-inductance $L_s$, and a screening parameter $\beta_L = 2 L_s I_c / \Phi_0$ where $I_c$ is the ideal critical current per JJ for zero spread and $\Phi_0$ is the flux quantum. Each individual junction has its own critical current $I_{c,k}$ and normal-state resistance $R_k$, where $k$ denotes the junction number from left to right. These parameters are statistically correlated, as has been found from experimental studies~\cite{GRO97, HIL02, HAL03, MIT16, JEN09}; and follow a power law of the form
\begin{equation}\label{eq:Gross}
r_k = i_{c,k}^{-1/2},
\end{equation}
where $r_k = R_k/R$ and $i_{c,k} = I_{c,k}/I_c$, and $R$ is the resistance per JJ for zero spread. In YBCO, the fabricated junctions have large variations in $I_{c,k}$ and $R_k$, as has been shown experimentally~\cite{SHA03, LAM14, DU14}. We denote the standard deviation (spread) from the mean $I_c$ as $\sigma$, and to sample physical values for individual critical currents in our simulations we use a log-normal probability distribution with mean 1~\cite{CRO87, NIE24b}
\begin{align}\label{eq:lognorm}
    \rho (i_{c,k}) = \frac{1}{i_{c,k}\gamma \sqrt{2\pi}} \;  \exp\left[ -\frac{1}{2}\left(\frac{\ln{i_{c,k}}-\mu}{\gamma}  \right)^2\right] \, , 
\end{align}
where $\mu=-\gamma^2/2$ and $\gamma = \sqrt{\ln(1 + \sigma^2)}$. For small $\sigma$, this distribution approaches a Gaussian. At a finite operating temperature $T$, each JJ exhibits Johnson noise~\cite{TES77, VOS81}. The thermal noise strength $\Gamma_k$ of the $k$th junction is $\Gamma_k = \Gamma / r_k$, where $\Gamma = 2\pi k_B T / \Phi_0 I_c$, and $k_B$ is the Boltzmann constant. 

To investigate the impact of spread $\sigma$ on the performance of these SQUID arrays, we introduce the dimensionless variable $\eta$, where:
\begin{equation}
    \eta (N_p, \beta_L,\Gamma, \sigma) = \frac{\langle \Delta\overline{v}\, (N_p,\beta_L,\Gamma, \sigma) \rangle}{ \Delta\overline{v}\, (N_p,\beta_L,\Gamma, 0)},
\end{equation}
which is the ratio of the maximum ensemble averaged modulation depth $\langle \Delta\bar{v}\, (N_p,\beta_L,\Gamma, \sigma) \rangle$ at a given $\sigma$, to the maximum modulation depth at $\sigma=0$ (e.g. an ideal array with no spread in $I_{c,k}$). Here $\Delta\overline{v}$ is the peak-to-peak voltage normalized by $RI_c$. 

We calculate $\langle \Delta\overline{v}\rangle$ via Monte Carlo simulation using 1000 independent realizations, following the method developed in~\cite{GAL22a}. In all simulations we use $I_c=20\; \upmu$A and $R=10\; \Omega$ which is typical for YBCO SQUID arrays~\cite{GAL21, GAL22a, GAL24}. Additionally, we use the following fixed values for film thickness $d=0.2\; \upmu$m, London penetration depth $\lambda_L = 0.4\; \upmu$m and temperature $T=77$ K in all simulations. It is also important to note that, for each realization, we determine the optimum normalized dc bias current $i_b^* = I_b^* / I_c$ which maximizes $\Delta\overline{v}$~\cite{Crete2019, Gali2023, Berggren2023, Berggren2024}. 

\section{Discussion}
\begin{figure}[h!]
	\centering
	\includegraphics[width=0.5\textwidth]{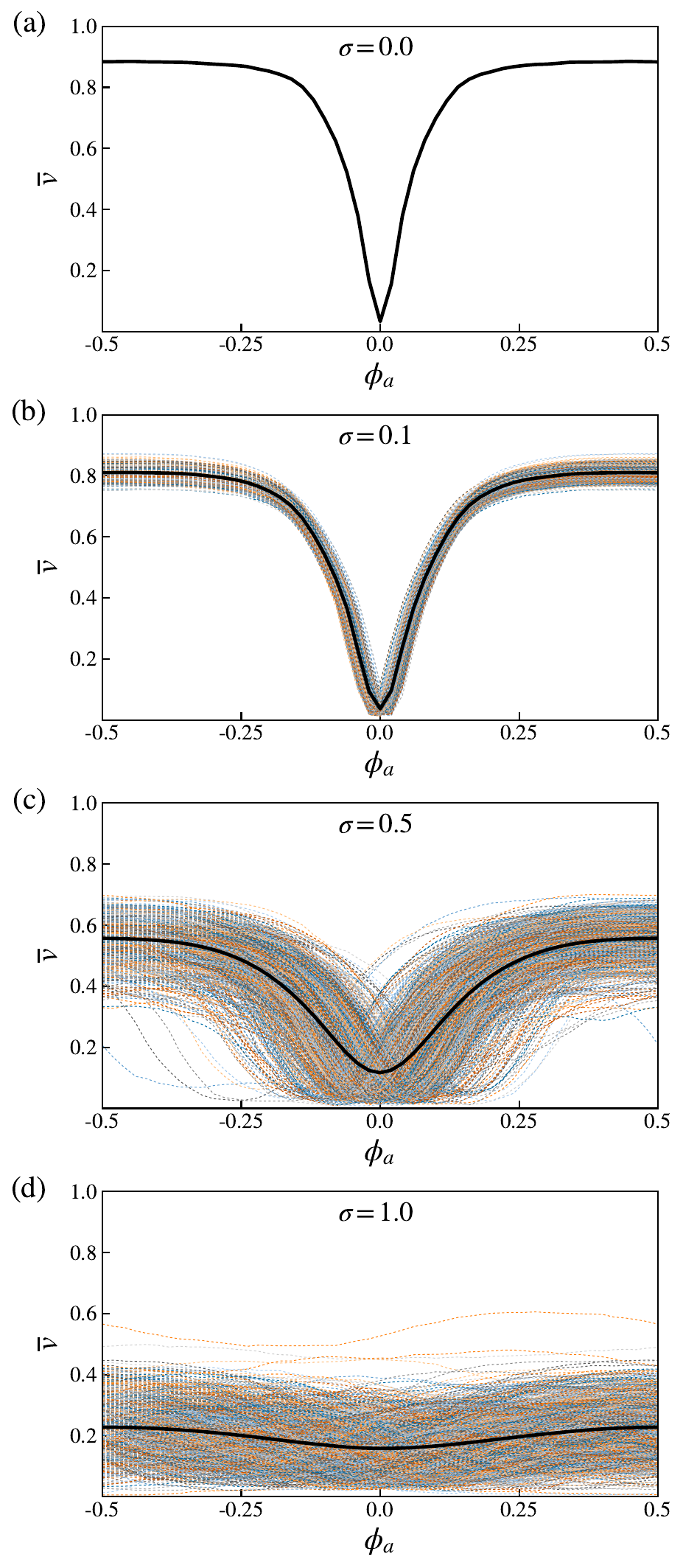}
	\caption{1000 independent Monte Carlo realizations (dashed colored lines) of the time-averaged voltage $\overline{v}$ versus normalized flux $\phi_a$ response of a SQUID array with $N_p=10$, using $\beta_L=1.0$ and $\Gamma=0.16$ as fixed parameters. The solid black line is the Monte Carlo ensemble average.}
	\label{fig:MCplots}
\end{figure}

The effect of $\sigma$ on the normalized averaged voltage-flux ($\overline{v}$ versus $\phi_a=\Phi_a / \Phi_0$, where $\Phi_a = B_a A_{\text{eff}}$ and $A_{\text{eff}}$ is the effective area of a SQUID cell) response is shown in Fig.~\ref{fig:MCplots}, for a SQUID array with $N_p=10$ and fixed values of $\beta_L=1.0$ and $\Gamma=0.16$. The fine dotted lines in Fig.~\ref{fig:MCplots} represent $\overline{v}$ versus $\phi_a$ for independent realizations calculated at each of the values $\sigma=0,\; 0.1,\; 0.5,\; 1.0$. The solid black lines represent the ensemble average $\langle \overline{v}\rangle$ versus $\phi_a$ for the 1000 Monte Carlo realizations calculated for each $\sigma$.
\begin{figure}[h!]
	\centering
	\includegraphics[width=0.5\textwidth]{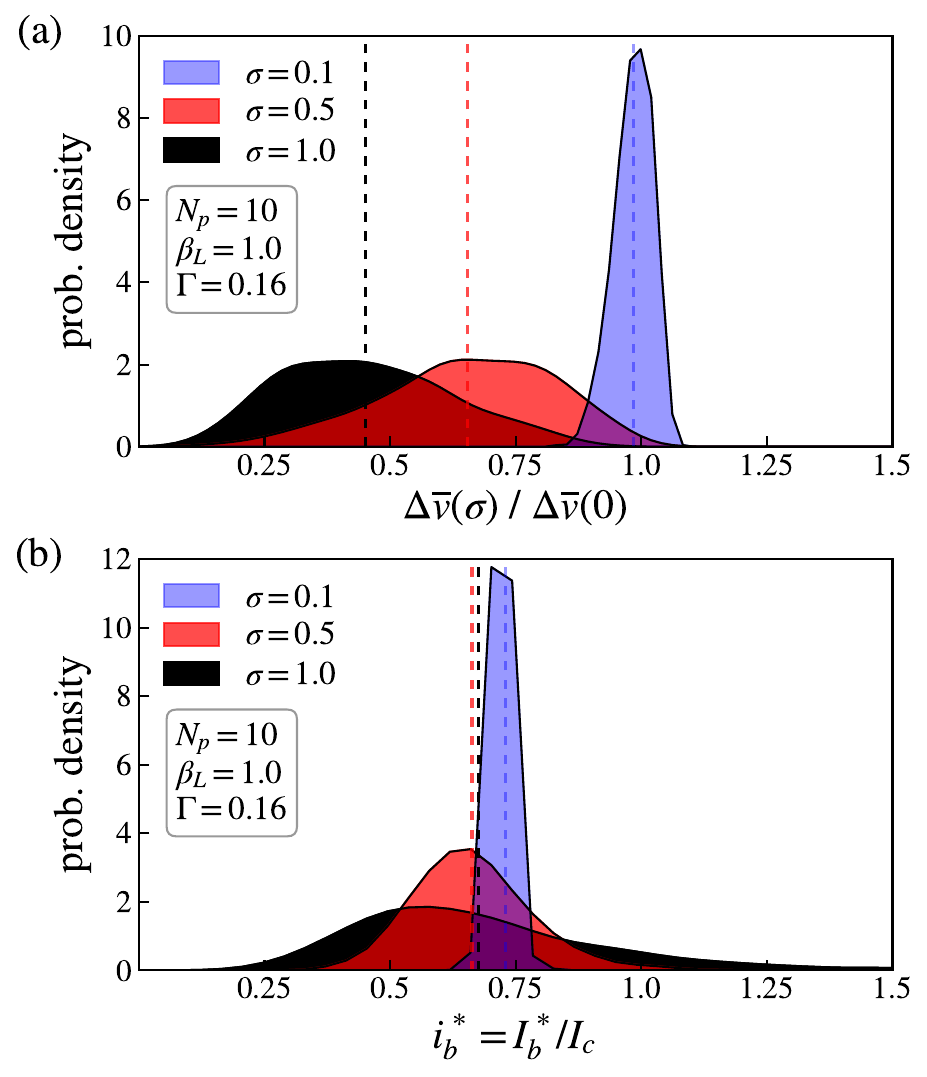}
	\caption{Probability density plots of (a) normalized peak-to-peak voltage and (b) optimum bias current for arrays with $N_p=10$, $\beta_L=1.0$ and $\Gamma=0.16$ and varying $\sigma$ values, using 1000 independent Monte Carlo realizations. The dashed vertical lines show the location of the corresponding mean values of each distribution, namely $\eta$ and $\langle i_b^*\rangle$ respectively.}
	\label{fig:ib-histogram}
\end{figure}

In the case of $\sigma=0$, all voltage-flux curves overlap, and the voltage-flux relationship has a sharp anti-peak centered at $\phi_a=0$. As $\sigma$ increases, the difference between independent Monte Carlo realizations becomes move obvious, and for $\sigma\geq 0.5$ most voltage-flux curves become asymmetric, with their anti-peaks off-centered. This shows that variations in junction parameters in the same array can produce vastly different responses to applied flux.

In Fig.~\ref{fig:ib-histogram} we see the results from plotting the distribution of $\Delta\overline{v}(\sigma)$ values relative to the zero spread case $\Delta\overline{v}(0)$. Here we see that for small $\sigma$, the distribution of $\Delta\overline{v}$ is approximately Gaussian (blue), while for larger $\sigma$ it becomes positively skewed, resembling a log-normal distribution (black). The optimum bias current $i_b^*$ which maximizes $\Delta\overline{v}$ in each case also follows a similar trend, however its mean value $\langle i_b^*\rangle$ does not change greatly with $\sigma$ and remains close to 0.75, corresponding to the optimum bias current when $\sigma=0$. Since $i_b^*(\sigma=0)$ depends on the device design and film properties~\cite{MUL21, MUL24}, the fact that it does not change significantly with $\sigma$ implies that one cannot accurately estimate $\sigma$ from $i_b^*$ in an experimental setting, and therefore we must rely on single JJ measurements to get an appropriate estimate for $\sigma$.

\begin{figure}[h!]
	\centering
	\includegraphics[width=0.5\textwidth]{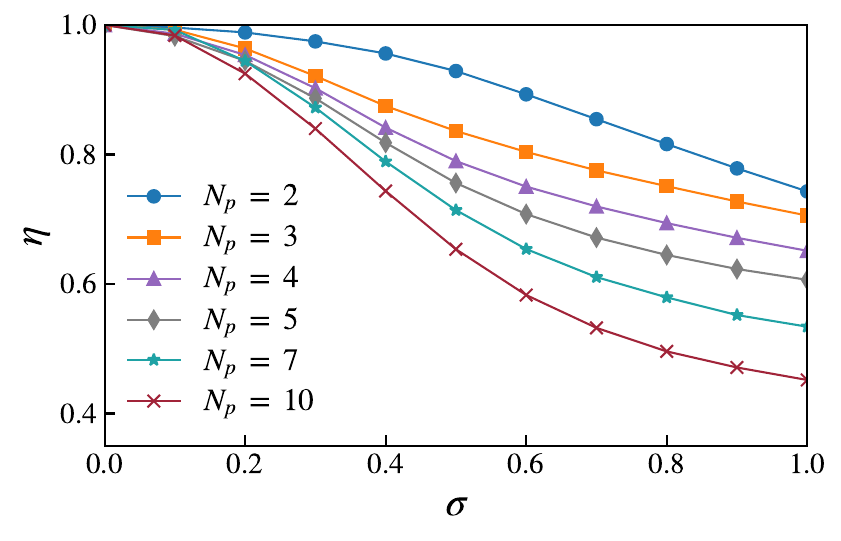}
	\caption{Normalized voltage modulation depth $\eta$ versus $\sigma$ of SQUID arrays with varying $N_p$. Here, we use a Monte Carlo ensemble average over 1000 independent realizations, with fixed parameters: $\beta_L=1.0$ and $\Gamma=0.16$.}
	\label{fig:1D-Np_MC}
\end{figure}

We now look at the variation in $\Delta\overline{v}$ with $N_p$. In Fig.~\ref{fig:1D-Np_MC}, $\eta$ decreases with $\sigma$ more rapidly as $N_p$ increases. This suggests that adding more parallel junctions into the array makes it more sensitive to junction spreads, and therefore one should limit $N_p$ when designing SQUID arrays for maximum performance. This is consistent with previous results with zero spread~\cite{KOR10, GAL21}, where the concept of a coupling radius was introduced and it was shown that the modulation depth decreases for $N_p > 6$. In our case, setting $\sigma > 0$ immediately results in a reduction in performance even for $N_p < 6$.
\begin{figure}[h!]
    \centering
    \includegraphics[width=0.5\textwidth]{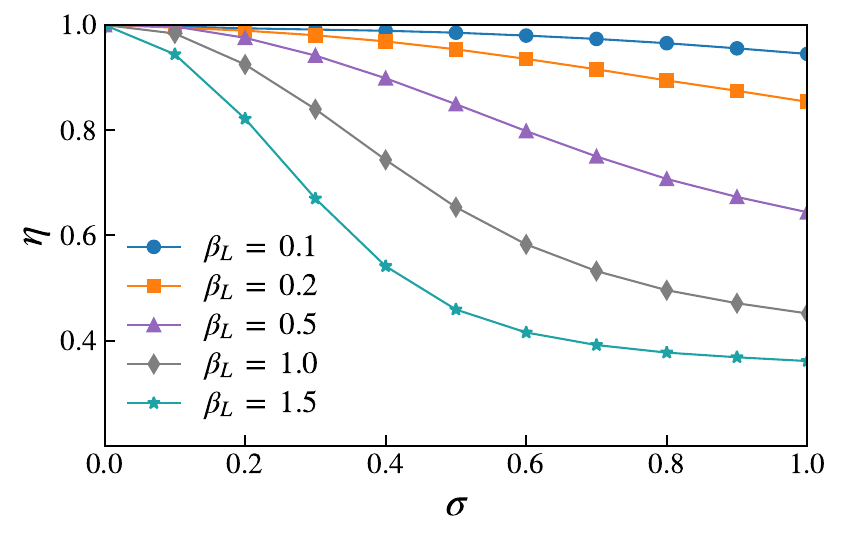}
    \caption{Normalized voltage modulation depth $\eta$ versus junction spread $\sigma$ of 1D arrays with $N_p=10$ for different screening parameters $\beta_L$ and fixed $\Gamma =0.16$.}
    \label{fig:eta-vs-sigma_betaL}
\end{figure}

Next, we look at the effect of varying the screening parameter $\beta_L=2L_s I_c/\Phi_0$. This is achieved by varying the SQUID cell size, which changes the cell inductance $L_s$. We fix the other parameters at $N_p=10$ and $\Gamma=0.16$. Figure~\ref{fig:eta-vs-sigma_betaL} shows that for small $\beta_L$, one gets a slowly decreasing $\eta$ with respect to $\sigma$, indicating that the array is more robust to junction parameter spreads in this regime. The reduction in $\eta$ is more pronounced with larger $\beta_L$, indicating that one should keep $\beta_L$ as small as possible in order to mitigate the effect of junction parameter spread.

\begin{figure}[h!]
	\centering
	\includegraphics[width=0.5\textwidth]{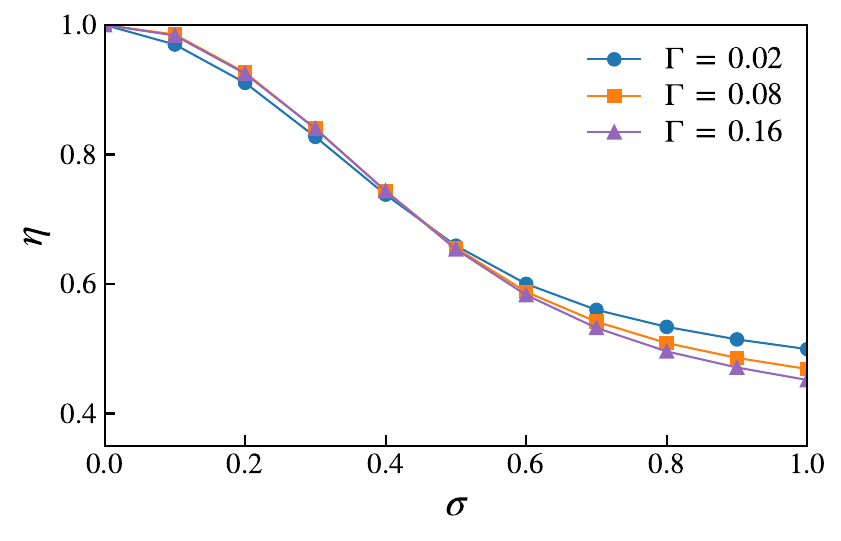}
	\caption{Normalized voltage modulation depth $\eta$ versus junction spread $\sigma$ of 1D arrays with $N_p=10$ for different noise strengths $\Gamma$ and fixed $\beta_L =1$.}
	\label{fig:eta-vs-sigma_Gamma}
\end{figure}

Lastly, we study the effect of $\Gamma$ in isolation from $N_p$ and $\beta_L$ by varying the temperature $T$, but keeping $I_c$ and $L_s$ constant. Unlike $N_p$ and $\beta_L$, $\Gamma$ has no significant impact on $\eta$, as shown in Fig.~\ref{fig:eta-vs-sigma_Gamma}. Although thermal noise is known to reduce $\Delta \overline{v}$~\cite{GAL22a, NIE24}, we see here that large $\sigma$ dominates the array's dc voltage response.

%__________________________________________________________________________________________________________________________
\section{Conclusion}\label{sec:Con}
In this paper, we analyzed the impact of junction parameter spread $\sigma$ on the peak-to-peak voltage versus flux response of one-dimensional commensurate SQUID arrays. Through Monte Carlo simulations, we showed that the average peak-to-peak voltage decreases rapidly with $\sigma$. This decrease is more prominent for larger $N_p$ and $\beta_L$. The impact of $\Gamma$ however was not significant, showing that the critical current disorder dominates the array's dc voltage response. Furthermore, keeping $\beta_L$ as small as possible can significantly mitigate the impact of $\sigma$. It would be interesting to expand this analysis to non-commensurate arrays (e.g. SQIF arrays) and in this case also include the effect of variation in loop-sizes, but due to the high computational complexity of these simulations, this will be be investigated in a separate work.

%____________________________________________________________________________
\section*{Author Contribution Statements}
{\bf M.A.G.L.}: Conceptualization, Investigation, Software, Visualization and Writing - Original Draft.
{\bf O.A.N.}: Formal Analysis, Investigation, Software, Visualization and Writing - Original Draft.
{\bf A.C.K.}: Formal Analysis, Investigation, Software, Validation and Writing - Review \& Editing.
{\bf K-H.M.}: Formal Analysis, Investigation, Validation and Writing - Review \& Editing.
{\bf E.E.M.}: Conceptualization, Experimental Design, Measurement - Review \& Editing.

% \appendix
% \section{Empirical formula for $\eta$}
% It may be of interest to the reader that we found an empirical formula for $\eta (\sigma; N_s, N_p, \beta_L, \Gamma)$ which behaves like a Lorentzian function with respect to $\sigma$:
% \begin{align}
%     \eta %(\sigma; N_s, N_p, \beta_L, \Gamma)
%     &\approx \frac{1 + i_b N_sN_p\beta_L \left( 1 + \left(p^2 \Gamma \right)^{1/p}\left(2\beta_L+ f(N_p, \sigma)  \right)^p\sigma^p \right)^{-1}}{1 + i_b N_sN_p\beta_L} \, , 
%     % B & \equiv (N_p N_s)^{\beta_L} \, ,\\
%     % C & \equiv 2\beta_L+N_p/10 \, , \text{works quite well for }Ns=1 \, ,\\
%     % C & \equiv 3.8\exp(-1/N_p) \, ,
% \end{align}
% where $p\approx 2.5$ based on the simulated data from Figures~\ref{fig:1D-Np_MC}$-$\ref{fig:eta-vs-sigma_Gamma}. 
% The function $f(N_p, \sigma)$ can be approximated to
% \begin{equation}
% f(N_p, \sigma) = \frac{N_p}{N_p^*} + \frac{1 - N_p/N_p^*}{1 + \exp(-N_p^*(N_p - N_p^*)} \, ,    
% \end{equation}
% where $N_p^*(\sigma)$ is the size of the coupling radius (maximum number of junctions in parallel that lead to an increase in the peak-to-peak voltage~\cite{GAL21}) at a given $\sigma$.
% For large arrays ($N_s \geq 10$) and for $\beta_L \sim 1$, the simple approximation $\eta \approx 1 - \sigma^2 $ gives us a good enough fit to estimate the strong decay of the array performance in the region $\sigma \leq 1$.

% \newpage
% Bibliography
%\section*{References}
% Bibliography: source file
\bibliography{main.bib}

\begin{thebibliography}{32}%
\makeatletter
\providecommand \@ifxundefined [1]{%
 \@ifx{#1\undefined}
}%
\providecommand \@ifnum [1]{%
 \ifnum #1\expandafter \@firstoftwo
 \else \expandafter \@secondoftwo
 \fi
}%
\providecommand \@ifx [1]{%
 \ifx #1\expandafter \@firstoftwo
 \else \expandafter \@secondoftwo
 \fi
}%
\providecommand \natexlab [1]{#1}%
\providecommand \enquote  [1]{``#1''}%
\providecommand \bibnamefont  [1]{#1}%
\providecommand \bibfnamefont [1]{#1}%
\providecommand \citenamefont [1]{#1}%
\providecommand \href@noop [0]{\@secondoftwo}%
\providecommand \href [0]{\begingroup \@sanitize@url \@href}%
\providecommand \@href[1]{\@@startlink{#1}\@@href}%
\providecommand \@@href[1]{\endgroup#1\@@endlink}%
\providecommand \@sanitize@url [0]{\catcode `\\12\catcode `\$12\catcode `\&12\catcode `\#12\catcode `\^12\catcode `\_12\catcode `\%12\relax}%
\providecommand \@@startlink[1]{}%
\providecommand \@@endlink[0]{}%
\providecommand \url  [0]{\begingroup\@sanitize@url \@url }%
\providecommand \@url [1]{\endgroup\@href {#1}{\urlprefix }}%
\providecommand \urlprefix  [0]{URL }%
\providecommand \Eprint [0]{\href }%
\providecommand \doibase [0]{http://dx.doi.org/}%
\providecommand \selectlanguage [0]{\@gobble}%
\providecommand \bibinfo  [0]{\@secondoftwo}%
\providecommand \bibfield  [0]{\@secondoftwo}%
\providecommand \translation [1]{[#1]}%
\providecommand \BibitemOpen [0]{}%
\providecommand \bibitemStop [0]{}%
\providecommand \bibitemNoStop [0]{.\EOS\space}%
\providecommand \EOS [0]{\spacefactor3000\relax}%
\providecommand \BibitemShut  [1]{\csname bibitem#1\endcsname}%
\let\auto@bib@innerbib\@empty
%</preamble>
\bibitem [{\citenamefont {Clarke}\ and\ \citenamefont {Braginski}(2006)}]{CLA06}%
  \BibitemOpen
  \bibfield  {author} {\bibinfo {author} {\bibfnamefont {J.}~\bibnamefont {Clarke}}\ and\ \bibinfo {author} {\bibfnamefont {A.}~\bibnamefont {Braginski}},\ }\href@noop {} {\emph {\bibinfo {title} {The SQUID handbook volume II: Applications of SQUIDs and SQUID systems}}}\ (\bibinfo  {publisher} {Wiley Online Library},\ \bibinfo {year} {2006})\BibitemShut {NoStop}%
\bibitem [{\citenamefont {Fagaly}(2006)}]{FAG06}%
  \BibitemOpen
  \bibfield  {author} {\bibinfo {author} {\bibfnamefont {R.}~\bibnamefont {Fagaly}},\ }\href@noop {} {\bibfield  {journal} {\bibinfo  {journal} {Review of scientific instruments}\ }\textbf {\bibinfo {volume} {77}} (\bibinfo {year} {2006})}\BibitemShut {NoStop}%
\bibitem [{\citenamefont {M{\"u}ller}\ and\ \citenamefont {Mitchell}(2021)}]{MUL21}%
  \BibitemOpen
  \bibfield  {author} {\bibinfo {author} {\bibfnamefont {K.-H.}\ \bibnamefont {M{\"u}ller}}\ and\ \bibinfo {author} {\bibfnamefont {E.}~\bibnamefont {Mitchell}},\ }\href@noop {} {\bibfield  {journal} {\bibinfo  {journal} {Physical Review B}\ }\textbf {\bibinfo {volume} {103}},\ \bibinfo {pages} {054509} (\bibinfo {year} {2021})}\BibitemShut {NoStop}%
\bibitem [{\citenamefont {Gal\'{\i}~Labarias}\ \emph {et~al.}(2022)\citenamefont {Gal\'{\i}~Labarias}, \citenamefont {M\"uller},\ and\ \citenamefont {Mitchell}}]{GAL22a}%
  \BibitemOpen
  \bibfield  {author} {\bibinfo {author} {\bibfnamefont {M.}~\bibnamefont {Gal\'{\i}~Labarias}}, \bibinfo {author} {\bibfnamefont {K.-H.}\ \bibnamefont {M\"uller}}, \ and\ \bibinfo {author} {\bibfnamefont {E.}~\bibnamefont {Mitchell}},\ }\href {\doibase 10.1103/PhysRevApplied.17.064009} {\bibfield  {journal} {\bibinfo  {journal} {Phys. Rev. Appl.}\ }\textbf {\bibinfo {volume} {17}},\ \bibinfo {pages} {064009} (\bibinfo {year} {2022})}\BibitemShut {NoStop}%
\bibitem [{\citenamefont {Labarias}\ \emph {et~al.}(2024)\citenamefont {Labarias}, \citenamefont {Nieves}, \citenamefont {Keenan},\ and\ \citenamefont {Mitchell}}]{GAL24}%
  \BibitemOpen
  \bibfield  {author} {\bibinfo {author} {\bibfnamefont {M.~A.~G.}\ \bibnamefont {Labarias}}, \bibinfo {author} {\bibfnamefont {O.~A.}\ \bibnamefont {Nieves}}, \bibinfo {author} {\bibfnamefont {S.~T.}\ \bibnamefont {Keenan}}, \ and\ \bibinfo {author} {\bibfnamefont {E.~E.}\ \bibnamefont {Mitchell}},\ }\href {\doibase 10.1109/TASC.2024.3396129} {\bibfield  {journal} {\bibinfo  {journal} {IEEE Transactions on Applied Superconductivity}\ }\textbf {\bibinfo {volume} {34}},\ \bibinfo {pages} {1} (\bibinfo {year} {2024})}\BibitemShut {NoStop}%
\bibitem [{\citenamefont {Nieves}\ and\ \citenamefont {Müller}(2024)}]{NIE24}%
  \BibitemOpen
  \bibfield  {author} {\bibinfo {author} {\bibfnamefont {O.~A.}\ \bibnamefont {Nieves}}\ and\ \bibinfo {author} {\bibfnamefont {K.-H.}\ \bibnamefont {Müller}},\ }\href {\doibase 10.1088/1361-6668/ad70dd} {\bibfield  {journal} {\bibinfo  {journal} {Superconductor Science and Technology}\ }\textbf {\bibinfo {volume} {37}},\ \bibinfo {pages} {105003} (\bibinfo {year} {2024})}\BibitemShut {NoStop}%
\bibitem [{\citenamefont {Jeng}\ \emph {et~al.}(2009)\citenamefont {Jeng}, \citenamefont {Lu}, \citenamefont {Wang},\ and\ \citenamefont {Wu}}]{JEN09}%
  \BibitemOpen
  \bibfield  {author} {\bibinfo {author} {\bibfnamefont {J.-T.}\ \bibnamefont {Jeng}}, \bibinfo {author} {\bibfnamefont {C.-C.}\ \bibnamefont {Lu}}, \bibinfo {author} {\bibfnamefont {C.-C.}\ \bibnamefont {Wang}}, \ and\ \bibinfo {author} {\bibfnamefont {C.-H.}\ \bibnamefont {Wu}},\ }\href@noop {} {\bibfield  {journal} {\bibinfo  {journal} {IEEE transactions on applied superconductivity}\ }\textbf {\bibinfo {volume} {19}},\ \bibinfo {pages} {214} (\bibinfo {year} {2009})}\BibitemShut {NoStop}%
\bibitem [{\citenamefont {Mitchell}\ and\ \citenamefont {Foley}(2010)}]{MIT10}%
  \BibitemOpen
  \bibfield  {author} {\bibinfo {author} {\bibfnamefont {E.~E.}\ \bibnamefont {Mitchell}}\ and\ \bibinfo {author} {\bibfnamefont {C.~P.}\ \bibnamefont {Foley}},\ }\href {\doibase 10.1088/0953-2048/23/6/065007} {\bibfield  {journal} {\bibinfo  {journal} {Superconductor Science and Technology}\ }\textbf {\bibinfo {volume} {23}},\ \bibinfo {pages} {065007} (\bibinfo {year} {2010})}\BibitemShut {NoStop}%
\bibitem [{\citenamefont {Lam}\ \emph {et~al.}(2014)\citenamefont {Lam}, \citenamefont {Lazar}, \citenamefont {Du},\ and\ \citenamefont {Foley}}]{LAM14}%
  \BibitemOpen
  \bibfield  {author} {\bibinfo {author} {\bibfnamefont {S.}~\bibnamefont {Lam}}, \bibinfo {author} {\bibfnamefont {J.}~\bibnamefont {Lazar}}, \bibinfo {author} {\bibfnamefont {J.}~\bibnamefont {Du}}, \ and\ \bibinfo {author} {\bibfnamefont {C.}~\bibnamefont {Foley}},\ }\href@noop {} {\bibfield  {journal} {\bibinfo  {journal} {Superconductor Science and Technology}\ }\textbf {\bibinfo {volume} {27}},\ \bibinfo {pages} {055011} (\bibinfo {year} {2014})}\BibitemShut {NoStop}%
\bibitem [{\citenamefont {Du}\ \emph {et~al.}(2014)\citenamefont {Du}, \citenamefont {Lazar}, \citenamefont {Lam}, \citenamefont {Mitchell},\ and\ \citenamefont {Foley}}]{DU14}%
  \BibitemOpen
  \bibfield  {author} {\bibinfo {author} {\bibfnamefont {J.}~\bibnamefont {Du}}, \bibinfo {author} {\bibfnamefont {J.}~\bibnamefont {Lazar}}, \bibinfo {author} {\bibfnamefont {S.}~\bibnamefont {Lam}}, \bibinfo {author} {\bibfnamefont {E.}~\bibnamefont {Mitchell}}, \ and\ \bibinfo {author} {\bibfnamefont {C.}~\bibnamefont {Foley}},\ }\href@noop {} {\bibfield  {journal} {\bibinfo  {journal} {Superconductor Science and Technology}\ }\textbf {\bibinfo {volume} {27}},\ \bibinfo {pages} {095005} (\bibinfo {year} {2014})}\BibitemShut {NoStop}%
\bibitem [{\citenamefont {Ohkubo}\ \emph {et~al.}(2022)\citenamefont {Ohkubo}, \citenamefont {Uehara}, \citenamefont {Beyer}, \citenamefont {Mimura}, \citenamefont {Tanaka}, \citenamefont {Ehara}, \citenamefont {Tanaka}, \citenamefont {Noguchi}, \citenamefont {Mitchell}, \citenamefont {Foley},\ and\ \citenamefont {Fagaly}}]{OHK22}%
  \BibitemOpen
  \bibfield  {author} {\bibinfo {author} {\bibfnamefont {M.}~\bibnamefont {Ohkubo}}, \bibinfo {author} {\bibfnamefont {G.}~\bibnamefont {Uehara}}, \bibinfo {author} {\bibfnamefont {J.}~\bibnamefont {Beyer}}, \bibinfo {author} {\bibfnamefont {M.}~\bibnamefont {Mimura}}, \bibinfo {author} {\bibfnamefont {H.}~\bibnamefont {Tanaka}}, \bibinfo {author} {\bibfnamefont {K.}~\bibnamefont {Ehara}}, \bibinfo {author} {\bibfnamefont {S.}~\bibnamefont {Tanaka}}, \bibinfo {author} {\bibfnamefont {T.}~\bibnamefont {Noguchi}}, \bibinfo {author} {\bibfnamefont {E.~E.}\ \bibnamefont {Mitchell}}, \bibinfo {author} {\bibfnamefont {C.~P.}\ \bibnamefont {Foley}}, \ and\ \bibinfo {author} {\bibfnamefont {R.~L.}\ \bibnamefont {Fagaly}},\ }\href {\doibase 10.1088/1361-6668/ac4f3b} {\bibfield  {journal} {\bibinfo  {journal} {Superconductor Science and Technology}\ }\textbf {\bibinfo {volume} {35}},\ \bibinfo {pages} {045002} (\bibinfo {year} {2022})}\BibitemShut {NoStop}%
\bibitem [{\citenamefont {Tesche}\ and\ \citenamefont {Clarke}(1977)}]{TES77}%
  \BibitemOpen
  \bibfield  {author} {\bibinfo {author} {\bibfnamefont {C.~D.}\ \bibnamefont {Tesche}}\ and\ \bibinfo {author} {\bibfnamefont {J.}~\bibnamefont {Clarke}},\ }\href@noop {} {\bibfield  {journal} {\bibinfo  {journal} {Journal of Low Temperature Physics}\ }\textbf {\bibinfo {volume} {29}},\ \bibinfo {pages} {301} (\bibinfo {year} {1977})}\BibitemShut {NoStop}%
\bibitem [{\citenamefont {Muller}\ \emph {et~al.}(2001)\citenamefont {Muller}, \citenamefont {Weiss}, \citenamefont {Gross}, \citenamefont {Kleiner},\ and\ \citenamefont {Koelle}}]{JMUL01}%
  \BibitemOpen
  \bibfield  {author} {\bibinfo {author} {\bibfnamefont {J.}~\bibnamefont {Muller}}, \bibinfo {author} {\bibfnamefont {S.}~\bibnamefont {Weiss}}, \bibinfo {author} {\bibfnamefont {R.}~\bibnamefont {Gross}}, \bibinfo {author} {\bibfnamefont {R.}~\bibnamefont {Kleiner}}, \ and\ \bibinfo {author} {\bibfnamefont {D.}~\bibnamefont {Koelle}},\ }\href@noop {} {\bibfield  {journal} {\bibinfo  {journal} {IEEE transactions on applied superconductivity}\ }\textbf {\bibinfo {volume} {11}},\ \bibinfo {pages} {912} (\bibinfo {year} {2001})}\BibitemShut {NoStop}%
\bibitem [{\citenamefont {Berggren}\ and\ \citenamefont {de~Escobar}(2014)}]{BER14}%
  \BibitemOpen
  \bibfield  {author} {\bibinfo {author} {\bibfnamefont {S.}~\bibnamefont {Berggren}}\ and\ \bibinfo {author} {\bibfnamefont {A.~L.}\ \bibnamefont {de~Escobar}},\ }\href@noop {} {\bibfield  {journal} {\bibinfo  {journal} {IEEE Transactions on Applied Superconductivity}\ }\textbf {\bibinfo {volume} {25}},\ \bibinfo {pages} {1} (\bibinfo {year} {2014})}\BibitemShut {NoStop}%
\bibitem [{\citenamefont {Labarias}\ and\ \citenamefont {Mitchell}(2023)}]{GAL23}%
  \BibitemOpen
  \bibfield  {author} {\bibinfo {author} {\bibfnamefont {M.~A.~G.}\ \bibnamefont {Labarias}}\ and\ \bibinfo {author} {\bibfnamefont {E.~E.}\ \bibnamefont {Mitchell}},\ }\href {\doibase 10.1109/TASC.2023.3236255} {\bibfield  {journal} {\bibinfo  {journal} {IEEE Transactions on Applied Superconductivity}\ }\textbf {\bibinfo {volume} {33}},\ \bibinfo {pages} {1} (\bibinfo {year} {2023})}\BibitemShut {NoStop}%
\bibitem [{\citenamefont {Wu}\ \emph {et~al.}(2012)\citenamefont {Wu}, \citenamefont {Cybart}, \citenamefont {Anton},\ and\ \citenamefont {Dynes}}]{WU12}%
  \BibitemOpen
  \bibfield  {author} {\bibinfo {author} {\bibfnamefont {S.}~\bibnamefont {Wu}}, \bibinfo {author} {\bibfnamefont {S.~A.}\ \bibnamefont {Cybart}}, \bibinfo {author} {\bibfnamefont {S.}~\bibnamefont {Anton}}, \ and\ \bibinfo {author} {\bibfnamefont {R.}~\bibnamefont {Dynes}},\ }\href@noop {} {\bibfield  {journal} {\bibinfo  {journal} {IEEE transactions on applied superconductivity}\ }\textbf {\bibinfo {volume} {23}},\ \bibinfo {pages} {1600104} (\bibinfo {year} {2012})}\BibitemShut {NoStop}%
\bibitem [{\citenamefont {M\"uller}\ and\ \citenamefont {Mitchell}(2024)}]{MUL24}%
  \BibitemOpen
  \bibfield  {author} {\bibinfo {author} {\bibfnamefont {K.-H.}\ \bibnamefont {M\"uller}}\ and\ \bibinfo {author} {\bibfnamefont {E.~E.}\ \bibnamefont {Mitchell}},\ }\href {\doibase 10.1103/PhysRevB.109.054507} {\bibfield  {journal} {\bibinfo  {journal} {Phys. Rev. B}\ }\textbf {\bibinfo {volume} {109}},\ \bibinfo {pages} {054507} (\bibinfo {year} {2024})}\BibitemShut {NoStop}%
\bibitem [{\citenamefont {Cybart}\ \emph {et~al.}(2012)\citenamefont {Cybart}, \citenamefont {Dalichaouch}, \citenamefont {Wu}, \citenamefont {Anton}, \citenamefont {Drisko}, \citenamefont {Parker}, \citenamefont {Harteneck},\ and\ \citenamefont {Dynes}}]{CYB12}%
  \BibitemOpen
  \bibfield  {author} {\bibinfo {author} {\bibfnamefont {S.~A.}\ \bibnamefont {Cybart}}, \bibinfo {author} {\bibfnamefont {T.}~\bibnamefont {Dalichaouch}}, \bibinfo {author} {\bibfnamefont {S.}~\bibnamefont {Wu}}, \bibinfo {author} {\bibfnamefont {S.}~\bibnamefont {Anton}}, \bibinfo {author} {\bibfnamefont {J.}~\bibnamefont {Drisko}}, \bibinfo {author} {\bibfnamefont {J.}~\bibnamefont {Parker}}, \bibinfo {author} {\bibfnamefont {B.}~\bibnamefont {Harteneck}}, \ and\ \bibinfo {author} {\bibfnamefont {R.}~\bibnamefont {Dynes}},\ }\href@noop {} {\bibfield  {journal} {\bibinfo  {journal} {Journal of Applied Physics}\ }\textbf {\bibinfo {volume} {112}} (\bibinfo {year} {2012})}\BibitemShut {NoStop}%
\bibitem [{\citenamefont {Gross}\ \emph {et~al.}(1997)\citenamefont {Gross}, \citenamefont {Alff}, \citenamefont {Beck}, \citenamefont {Proehlich}, \citenamefont {Koelle},\ and\ \citenamefont {Marx}}]{GRO97}%
  \BibitemOpen
  \bibfield  {author} {\bibinfo {author} {\bibfnamefont {R.}~\bibnamefont {Gross}}, \bibinfo {author} {\bibfnamefont {L.}~\bibnamefont {Alff}}, \bibinfo {author} {\bibfnamefont {A.}~\bibnamefont {Beck}}, \bibinfo {author} {\bibfnamefont {M.}~\bibnamefont {Proehlich}}, \bibinfo {author} {\bibfnamefont {D.}~\bibnamefont {Koelle}}, \ and\ \bibinfo {author} {\bibfnamefont {A.}~\bibnamefont {Marx}},\ }\href {\doibase 10.1109/77.621919} {\bibfield  {journal} {\bibinfo  {journal} {IEEE Transactions on Applied Superconductivity}\ }\textbf {\bibinfo {volume} {7}},\ \bibinfo {pages} {2929} (\bibinfo {year} {1997})}\BibitemShut {NoStop}%
\bibitem [{\citenamefont {Hilgenkamp}\ and\ \citenamefont {Mannhart}(2002)}]{HIL02}%
  \BibitemOpen
  \bibfield  {author} {\bibinfo {author} {\bibfnamefont {H.}~\bibnamefont {Hilgenkamp}}\ and\ \bibinfo {author} {\bibfnamefont {J.}~\bibnamefont {Mannhart}},\ }\href@noop {} {\bibfield  {journal} {\bibinfo  {journal} {Reviews of Modern Physics}\ }\textbf {\bibinfo {volume} {74}},\ \bibinfo {pages} {485} (\bibinfo {year} {2002})}\BibitemShut {NoStop}%
\bibitem [{\citenamefont {Halbritter}(2003)}]{HAL03}%
  \BibitemOpen
  \bibfield  {author} {\bibinfo {author} {\bibfnamefont {J.}~\bibnamefont {Halbritter}},\ }\href@noop {} {\bibfield  {journal} {\bibinfo  {journal} {IEEE transactions on applied superconductivity}\ }\textbf {\bibinfo {volume} {13}},\ \bibinfo {pages} {1158} (\bibinfo {year} {2003})}\BibitemShut {NoStop}%
\bibitem [{\citenamefont {Mitchell}\ \emph {et~al.}(2016)\citenamefont {Mitchell}, \citenamefont {Hannam}, \citenamefont {Lazar}, \citenamefont {Leslie}, \citenamefont {Lewis}, \citenamefont {Grancea}, \citenamefont {Keenan}, \citenamefont {Lam},\ and\ \citenamefont {Foley}}]{MIT16}%
  \BibitemOpen
  \bibfield  {author} {\bibinfo {author} {\bibfnamefont {E.}~\bibnamefont {Mitchell}}, \bibinfo {author} {\bibfnamefont {K.}~\bibnamefont {Hannam}}, \bibinfo {author} {\bibfnamefont {J.}~\bibnamefont {Lazar}}, \bibinfo {author} {\bibfnamefont {K.}~\bibnamefont {Leslie}}, \bibinfo {author} {\bibfnamefont {C.}~\bibnamefont {Lewis}}, \bibinfo {author} {\bibfnamefont {A.}~\bibnamefont {Grancea}}, \bibinfo {author} {\bibfnamefont {S.}~\bibnamefont {Keenan}}, \bibinfo {author} {\bibfnamefont {S.}~\bibnamefont {Lam}}, \ and\ \bibinfo {author} {\bibfnamefont {C.}~\bibnamefont {Foley}},\ }\href@noop {} {\bibfield  {journal} {\bibinfo  {journal} {Superconductor Science and Technology}\ }\textbf {\bibinfo {volume} {29}},\ \bibinfo {pages} {06LT01} (\bibinfo {year} {2016})}\BibitemShut {NoStop}%
\bibitem [{\citenamefont {Shadrin}\ \emph {et~al.}(2003)\citenamefont {Shadrin}, \citenamefont {Jia},\ and\ \citenamefont {Divin}}]{SHA03}%
  \BibitemOpen
  \bibfield  {author} {\bibinfo {author} {\bibfnamefont {P.}~\bibnamefont {Shadrin}}, \bibinfo {author} {\bibfnamefont {C.}~\bibnamefont {Jia}}, \ and\ \bibinfo {author} {\bibfnamefont {Y.}~\bibnamefont {Divin}},\ }\href@noop {} {\bibfield  {journal} {\bibinfo  {journal} {IEEE Transactions on Applied Superconductivity}\ }\textbf {\bibinfo {volume} {13}},\ \bibinfo {pages} {603} (\bibinfo {year} {2003})}\BibitemShut {NoStop}%
\bibitem [{\citenamefont {Crow}\ and\ \citenamefont {Shimizu}(1987)}]{CRO87}%
  \BibitemOpen
  \bibfield  {author} {\bibinfo {author} {\bibfnamefont {E.~L.}\ \bibnamefont {Crow}}\ and\ \bibinfo {author} {\bibfnamefont {K.}~\bibnamefont {Shimizu}},\ }\href@noop {} {\emph {\bibinfo {title} {Lognormal distributions}}}\ (\bibinfo  {publisher} {Marcel Dekker New York},\ \bibinfo {year} {1987})\BibitemShut {NoStop}%
\bibitem [{\citenamefont {Nieves}(2024)}]{NIE24b}%
  \BibitemOpen
  \bibfield  {author} {\bibinfo {author} {\bibfnamefont {O.~A.}\ \bibnamefont {Nieves}},\ }\href@noop {} {\emph {\bibinfo {title} {Gaussian Integrals and their Applications}}}\ (\bibinfo  {publisher} {CRC Press},\ \bibinfo {year} {2024})\BibitemShut {NoStop}%
\bibitem [{\citenamefont {Voss}(1981)}]{VOS81}%
  \BibitemOpen
  \bibfield  {author} {\bibinfo {author} {\bibfnamefont {R.~F.}\ \bibnamefont {Voss}},\ }\href {\doibase 10.1007/BF00116701} {\bibfield  {journal} {\bibinfo  {journal} {Low Temperature Physics}\ }\textbf {\bibinfo {volume} {42}},\ \bibinfo {pages} {151} (\bibinfo {year} {1981})}\BibitemShut {NoStop}%
\bibitem [{\citenamefont {Labarias}\ \emph {et~al.}(2021)\citenamefont {Labarias}, \citenamefont {M{\"u}ller},\ and\ \citenamefont {Mitchell}}]{GAL21}%
  \BibitemOpen
  \bibfield  {author} {\bibinfo {author} {\bibfnamefont {M.~G.}\ \bibnamefont {Labarias}}, \bibinfo {author} {\bibfnamefont {K.-H.}\ \bibnamefont {M{\"u}ller}}, \ and\ \bibinfo {author} {\bibfnamefont {E.}~\bibnamefont {Mitchell}},\ }\href@noop {} {\bibfield  {journal} {\bibinfo  {journal} {IEEE Transactions on Applied Superconductivity}\ }\textbf {\bibinfo {volume} {32}},\ \bibinfo {pages} {1} (\bibinfo {year} {2021})}\BibitemShut {NoStop}%
\bibitem [{\citenamefont {Crété}\ \emph {et~al.}(2019)\citenamefont {Crété}, \citenamefont {Lemaître}, \citenamefont {Marcilhac}, \citenamefont {Trastoy},\ and\ \citenamefont {Ulysse}}]{Crete2019}%
  \BibitemOpen
  \bibfield  {author} {\bibinfo {author} {\bibfnamefont {D.}~\bibnamefont {Crété}}, \bibinfo {author} {\bibfnamefont {Y.}~\bibnamefont {Lemaître}}, \bibinfo {author} {\bibfnamefont {B.}~\bibnamefont {Marcilhac}}, \bibinfo {author} {\bibfnamefont {J.}~\bibnamefont {Trastoy}}, \ and\ \bibinfo {author} {\bibfnamefont {C.}~\bibnamefont {Ulysse}},\ }in\ \href {\doibase 10.1109/ISEC46533.2019.8990895} {\emph {\bibinfo {booktitle} {2019 IEEE International Superconductive Electronics Conference (ISEC)}}}\ (\bibinfo {year} {2019})\ pp.\ \bibinfo {pages} {1--3}\BibitemShut {NoStop}%
\bibitem [{\citenamefont {Labarias}\ \emph {et~al.}(2023)\citenamefont {Labarias}, \citenamefont {Müller},\ and\ \citenamefont {Mitchell}}]{Gali2023}%
  \BibitemOpen
  \bibfield  {author} {\bibinfo {author} {\bibfnamefont {M.~A.~G.}\ \bibnamefont {Labarias}}, \bibinfo {author} {\bibfnamefont {K.-H.}\ \bibnamefont {Müller}}, \ and\ \bibinfo {author} {\bibfnamefont {E.~E.}\ \bibnamefont {Mitchell}},\ }\href {\doibase 10.1088/1361-6668/acfa7a} {\bibfield  {journal} {\bibinfo  {journal} {Superconductor Science and Technology}\ }\textbf {\bibinfo {volume} {36}},\ \bibinfo {pages} {115016} (\bibinfo {year} {2023})}\BibitemShut {NoStop}%
\bibitem [{\citenamefont {Berggren}\ \emph {et~al.}(2023)\citenamefont {Berggren}, \citenamefont {Crowe}, \citenamefont {Ferrante},\ and\ \citenamefont {Taylor}}]{Berggren2023}%
  \BibitemOpen
  \bibfield  {author} {\bibinfo {author} {\bibfnamefont {S.~A.~E.}\ \bibnamefont {Berggren}}, \bibinfo {author} {\bibfnamefont {S.~T.}\ \bibnamefont {Crowe}}, \bibinfo {author} {\bibfnamefont {N.~B.}\ \bibnamefont {Ferrante}}, \ and\ \bibinfo {author} {\bibfnamefont {B.~J.}\ \bibnamefont {Taylor}},\ }\href {\doibase 10.1109/TASC.2023.3258367} {\bibfield  {journal} {\bibinfo  {journal} {IEEE Transactions on Applied Superconductivity}\ }\textbf {\bibinfo {volume} {33}},\ \bibinfo {pages} {1} (\bibinfo {year} {2023})}\BibitemShut {NoStop}%
\bibitem [{\citenamefont {Berggren}\ \emph {et~al.}(2024)\citenamefont {Berggren}, \citenamefont {Crowe}, \citenamefont {Greenough}, \citenamefont {Sanborn}, \citenamefont {Ferrante},\ and\ \citenamefont {Taylor}}]{Berggren2024}%
  \BibitemOpen
  \bibfield  {author} {\bibinfo {author} {\bibfnamefont {S.~A.~E.}\ \bibnamefont {Berggren}}, \bibinfo {author} {\bibfnamefont {S.~T.}\ \bibnamefont {Crowe}}, \bibinfo {author} {\bibfnamefont {R.~D.}\ \bibnamefont {Greenough}}, \bibinfo {author} {\bibfnamefont {G.~P.}\ \bibnamefont {Sanborn}}, \bibinfo {author} {\bibfnamefont {N.~B.}\ \bibnamefont {Ferrante}}, \ and\ \bibinfo {author} {\bibfnamefont {B.~J.}\ \bibnamefont {Taylor}},\ }\href {\doibase 10.1109/TASC.2024.3367807} {\bibfield  {journal} {\bibinfo  {journal} {IEEE Transactions on Applied Superconductivity}\ }\textbf {\bibinfo {volume} {34}},\ \bibinfo {pages} {1} (\bibinfo {year} {2024})}\BibitemShut {NoStop}%
\bibitem [{\citenamefont {Kornev}\ \emph {et~al.}(2010)\citenamefont {Kornev}, \citenamefont {Soloviev}, \citenamefont {Klenov},\ and\ \citenamefont {Mukhanov}}]{KOR10}%
  \BibitemOpen
  \bibfield  {author} {\bibinfo {author} {\bibfnamefont {V.}~\bibnamefont {Kornev}}, \bibinfo {author} {\bibfnamefont {I.}~\bibnamefont {Soloviev}}, \bibinfo {author} {\bibfnamefont {N.}~\bibnamefont {Klenov}}, \ and\ \bibinfo {author} {\bibfnamefont {O.}~\bibnamefont {Mukhanov}},\ }\href@noop {} {\bibfield  {journal} {\bibinfo  {journal} {IEEE transactions on applied superconductivity}\ }\textbf {\bibinfo {volume} {21}},\ \bibinfo {pages} {394} (\bibinfo {year} {2010})}\BibitemShut {NoStop}%
\end{thebibliography}%

\end{document}